\documentclass{ws-ijmpa}
\usepackage[super,compress]{cite}
\usepackage{graphicx}
\begin{document}
\markboth{Pham Quang Hung and Nguyen Nhu Le}{Dynamical Electroweak Symmetry Breaking in the model of electroweak-scale right-handed neutrinos}

%
\catchline{}{}{}{}{}
%

\title{DYNAMICAL ELECTROWEAK SYMMETRY BREAKING IN THE MODEL OF ELECTROWEAK-SCALE RIGHT-HANDED NEUTRINOS
}

\author{PHAM QUANG HUNG
}

\address{Department of Physics, University of Virginia,\\
Charlottesville, VA 22904-4714, USA\\Center for Theoretical and Computational Physics, College of Education, Hue University\\Hue City, Vietnam\\
pqh@viginia.edu}

\author{Nguyen Nhu Le}

\address{Department of Physics, College of Education, Hue University, \\
Hue City, Vietnam\\Center for Theoretical and Computational Physics, College of Education, Hue University\\Hue City, Vietnam\\
naspears13@gmail.com}

\maketitle

\begin{history}
\received{Day Month Year}
\revised{Day Month Year}
\end{history}

\begin{abstract}
We present the Higgs mechanism in the context of the EW-scale $\nu_R$ model in which electroweak symmetry is dynamically broken by condensates of mirror quark and right-handed neutrino through the exchange of one fundamental Higgs doublet and one fundamental Higgs triplet, respectively. The formation of these condensates is dynamically investigated by using the Schwinger-Dyson approach. The occurrence of these condensates will give rise to the rich Higgs spectrum. In addition, the VEVs of Higgs fields is also discussed in this dynamical phenomenon.

\keywords{Dynamical Electroweak Symmetry Breaking; Condensate; EW-scale $\nu_R$.}
\end{abstract}

\ccode{PACS numbers:}

\tableofcontents

\section{Introduction}
It is generally agreed that the Standard Model (SM) is an incomplete theory since there remains many questions which necessitates a framework that goes beyond the SM (BSM). The origin of the (tiny) neutrino masses is one of them.\par
There have been many  "explanations" for the smallness of neutrino masses. The most popular mechanism is the see-saw mechanism \cite{Seesaw} in which light neutrino masses take the form $m_D^2/M_R$, where $m_D$ is the Dirac mass coming from a lepton-number conserving term and $M_R$ is the Majorana mass coming from a lepton number violating term. The nature of these two mass scales is very model-dependent. In the simplest version of see-saw mechanism, right-handed neutrinos are SM singlets and this Majorana mass term is huge compared with the Dirac mass term (which is proportional to the electroweak scale) in order to preserve a tiny mass $m_D^2/M_R$ for the lighter of the two eigenstates. It goes without saying that the right-handed neutrinos in this scenario are sterile and practically undetectable. What if right-handed neutrinos are non-sterile? Ref.~\cite{HungMH} proposed just such a model called the $EW-\nu_R$ model  in which right-handed neutrinos are non-sterile and have masses proportional to the electroweak scale $\Lambda_{EW} \sim 246~GeV$. In the EW-scale $\nu_R$ model, it is possible to detect right-handed neutrinos. 

The first version of EW-scale $\nu_R$ model \cite{HungMH} was published in 2007 in which the focus of interest is an explanation of the tiny mass of neutrino with right-handed neutrino at the electroweak scale. In the following years, some important issues of the model such as lepton flavor violating processes, consequences of Pati-Salam unification of active right-handed neutrino, Higgs contents are discussed in \cite{LepPaHi1, LepPaHi2, LepPaHi3}. This model also satisfy the constraints from the precision electroweak data \cite{VinhHungA} and its extended version \cite{Th.Hung} agrees with the experiment data of the 125-GeV SM-like Higgs boson. This undoubtedly affirms the viability of this model. The implications for the extended model for neutrino masses and mixings, lepton flavor violating decays and the search for mirror quarks at the LHC (Large Hadron Collider) were recently discussed \cite{neulepmirror1, neulepmirror2, neulepmirror3}.\par 
In this manuscript, we study the Dynamical Symmetry Breaking of the EW-scale $\nu_R$ model. The SM is spontaneously broken by the Higgs potential of the form $V(\phi)=-\mu^2\phi^+\phi+\lambda\left(\phi^+\phi\right)^2$, where $\phi$ is an elementary scalar field. This leaves many often-asked questions such as: why $\mu^2$ is positive, or the hierarchy problem: why the electroweak scale $v$ is smaller than the Plank scale $M_P$ by many orders of magnitude.  The most popular way to deal with this problem is to use the cancellation between the quadratically-divergent contributions of fermion and that of boson proposed in some interesting models come along with the Supersymmetry (SUSY), Little Higgs, Twin Higgs, etc... \cite{models}. Another idea can be found in the idea of Large Extra Dimensions, Higgsless models \cite{models} where the extra dimensions play an essential rule to avoid the hierarchy problem. Another class of models that does not involve elementary scalar fields is one in which symmetry breaking is realized dynamically through condensates of bilinear fermion fields. There are many models of this type such as: composite Higgs models, Technicolor (TC), Extended TC, top-color, etc...\cite{models}. In this work, we will present a scenario in which the EW-scale $\nu_R$ model is dynamically broken by a condensate. This scenario is based on the presented in Ref.~\cite{HungXiong}. In Ref.~\cite{HungXiong}, Dynamical Electroweak Symmetry Breaking (DEWSB) was accomplished by condensates on the one of
the 4th generation fermions through the exchange of a fundamental Higgs
doublet that has no $\mu^2\phi^+\phi$ term. In particular, this fundamental Higgs doublet is assumed to be  
massless and has no vacuum expectation value (VEV) at tree level. The model is assumed to be scale-invariant at tree-level.\par 
The Schwinger-Dyson equation was written down for the fermion self-energy. It was shown that when $\alpha_Y=g_Y^2/4\pi\geq\pi/2$, the solution to the SD equation admits a condensate solution. Condensate of the form $\langle\bar t_L't_R'\rangle$ can be written in terms of the $t'$ self-energy. From this scenario, it would be possible to define a physical cutoff scale of $\mathcal O(\mbox{TeV})$. How to get the condensate states and the Higgs mechanism will be the aims of this paper.

The organization of the paper will be as follows. A brief introduction to the EW-scale $\nu_R$ model will first be given. By using the Schwinger-Dyson (SD) approach, we then obtain the conditions under which the condensates get formed in the Higgs-Yukawa system. We next calculate the evolution of the Yukawa couplings at the one-loop level and constraint their initial values so that the condition for condensate formation occurs at an energy scale of $\mathcal O(1~ TeV)$ which is close to the location of the Landau pole. We then discuss issues of DEWSB related to the mechanism of mass generation. Implications of our
results concerning the nature of the light neutrino mass are presented in next section. We end with the discussion our results and conclusions.
\section{The EW-scale $\nu_R$ model in a nutshell}
Here we will only give a brief introduction to the EW-scale $\nu_R$ model. The motivation for constructing of \cite{HungMH} was to give an explanation for one of the important questions in particle physics: why neutrinos have tiny masses. One crucial question is the following: How do we test experimentally a model which can provide a natural explanation for the smallness of neutrino masses? The EW-scale $\nu_R$ model \cite{HungMH} was constructed with such aim in mind. As we have mentioned in the Introduction, a crucial test of the seesaw mechanism would be the detection of right-handed neutrinos. This is possible in the EW-scale $\nu_R$ model because right-handed neutrinos are parts of $SU(2)$ doublets along with their charged mirror leptons. The gauge group of the EW-scale $\nu_R$ model is $SU(3)_c \times SU(2)_W \times U(1)_Y$. More details can be found in \cite{HungMH}. We only present here fermion fields, Higgs fields and their interactions.   
\subsection{Fermion contents}
In addition to the SM particle content the EW-scale $\nu_R$ model \cite{HungMH} contains the supplemental fields
shown in Table~\ref{table11}
\hspace{2cm}
\begin{table}[ph]
\tbl{Fermion fields in the EW-scale $\nu_R$ model.}
{\begin{tabular}{@{}cccc@{}} \toprule
\bf{SM fermion fields } &\bf{$SU(2)_W\otimes U(1)_Y$} & \bf{Mirror fermion fields }  &
\bf{$SU(2)_W\otimes U(1)_Y$} \\
\colrule
$l_L=\left(\begin{matrix}
\nu_L\\e_L
\end{matrix}\right)$  & $\left(2,\displaystyle -\frac 12\right) $ & $l_R^M=\left(\begin{matrix}
\nu_R\\e_R^M
\end{matrix}\right)$  & $\left(2,\displaystyle -\frac 12\right) $   \\
 $e_R$ & $\left(1,-1\right)$ &  $e_L^M$&$\left(1,-1\right)$ \\
$q_L=\left(\begin{matrix}
u_L\\d_L
\end{matrix}\right)$  & $\left(2,\displaystyle \frac 16\right) $ &$q_R^M=\left(\begin{matrix}
u_R^M\\d_R^M
\end{matrix}\right)$&$\left(2,\displaystyle \frac 16\right) $ \\
$u_R$ & $\displaystyle\left(1,\frac 2 3\right)$ &$u_L^M$ & $\displaystyle\left(1,\frac 2 3\right)$  \\ 
$d_R$ & $\displaystyle\left(1,-\frac 1 3\right)$ & $d_L^M$ & $\displaystyle\left(1,-\frac 1 3\right)$  \\
\botrule
\end{tabular} \label{table11}}
\end{table}
\subsection{Higgs contents}
As it is discussed in \cite{HungMH}, the fermion bilinear $l_R^{M,T}\sigma_2l_R^M$ giving rise to a Majorana mass term for the right-handed neutrinos. A singlet Higgs field is ruled out because its non-zero VEV would break charge conservation \cite{HungMH}. Hence, a triplet Higgs is an appropriate choice. Nevertheless, if there exists only one Higgs triplet then the custodial symmetry, $\rho=1$, can not be preserved at tree level. Ref.~\cite{HungMH} introduced an additional Higgs triplet. To generate a neutrino Dirac mass term, a singlet Higgs field $\phi_S$ was introduced \cite{HungMH}. SM quark and charged lepton masses are obtained by a coupling to a Higgs doublet $\Phi_2$ \cite{HungMH} and those of mirror quarks and charged leptons come from a coupling to a second Higgs doublet $\Phi_{2M}$ \cite{Th.Hung}. The latter was needed \cite{Th.Hung} in order to accommodate the discovery of the 125-GeV scalar at the LHC \cite{126Higgs}. Higgs fields transforming under $SU(3)_c \times SU(2)_W \times U(1)_Y$ are listed as follows.  
\begin{itemize}
\item Higgs triplets
\begin{eqnarray}\label{HTrip}\widetilde\chi&=&\frac 1 {\sqrt 2}\vec\tau\cdot\vec\chi= \begin{pmatrix}
\frac 1 {\sqrt 2}\chi^+& \chi^{++}\\
{\chi^{0}}& -\frac 1 {\sqrt 2}\chi^+
\end{pmatrix}=\left(1,3,Y/2=1\right),\\\label{Higgsxi}\xi&=&\begin{pmatrix}\xi^+\\\xi^0\\\xi^-\end{pmatrix}=\left(1,3,Y/2=0\right).\end{eqnarray}
Higgs triplets such as in Eq.~\eqref{Higgsxi} were considered earlier in \cite{16Patisalam}.
\begin{equation}\label{10hh}\chi=\left(\begin{matrix}\chi^0&\xi^+&\chi^{++}\\\chi^-&\xi^0&\chi^+\\\chi^{--}&\xi^-&\chi^{0*}\end{matrix}\right).
\end{equation}
\item Higgs doublets 
\begin{eqnarray}\label{phi1}\Phi_{2}&=&\left(\begin{matrix}\phi_2^+\\\phi_2^0\end{matrix}\right)=\left(1,2,Y/2=1/2\right),\\\label{phi2}\Phi_{2M}&=&\left(\begin{matrix}\phi_{2M}^+\\\phi_{2M}^0\end{matrix}\right)=\left(1,2,Y/2=1/2\right).\end{eqnarray}
\item Higgs singlet
\begin{eqnarray}\label{dontuyen}\phi_S=(1,1,Y/2=0).\end{eqnarray}
\end{itemize}
\subsection{ Yukawa interactions}The interaction between Higgs and fermion fields in the EW-scale $\nu_R$ model can be found in \cite{HungMH} and \cite{LepPaHi3,VinhHungA,Th.Hung}. In the case of SM fermions, we have the usual generic Yukawa interactions for quarks and leptons.
\begin{eqnarray}\mathcal L_{YSM}=-g_{ij}\bar\Psi_{Li}\Phi_2\Psi_{Rj}+\mbox{h.c.}.\end{eqnarray}
For mirror fermion we need to consider the terms
 \begin{eqnarray}\label{mm1}\mathcal L_{e^M}&=&-g_{e^M}\bar l_R^M\Phi_{2M} e_L^M+\mbox{h.c}.,\\\label{mm2}\mathcal L_{q^M}&=&-g_{d^M}\bar q_R^M\Phi_{2M} d_L^M-g_{u^M}\bar q_R^M\widetilde\Phi_{2M} u_L^M+\mbox{h.c.},\\\label{mm3}\mathcal L_{\nu_R}&=&g_Ml_{R}^{M,T}\sigma_2\tau_2\widetilde\chi l_R^M,\\\label{dontuyenY1}\mathcal {L}_{Sl}&=&-g_{Sl}\bar l_Ll_R^M\phi_S+\mbox{h.c.},\\\label{dontuyenY2}\mathcal {L}_{Sq}&=&-g_{Sq}\bar q_R^Mq_L\phi_S-g_{Sq}'\bar q_L^Mq_R\phi_S+\mbox{h.c.}.\end{eqnarray}
Notice that, by imposing a global $U(1)$ symmetry on the fermion and scalar sectors, the only allowed Yukawa couplings to the singlet scalar $\phi_S$ are shown in Eqs. \eqref{dontuyenY1} and \eqref{dontuyenY2}. The details of this global symmetry can be found in Ref. \cite{HungMH} and Ref. \cite{Th.Hung}. The interactions given by Eqs.~\eqref{mm3} and \eqref{dontuyenY1} would provide Majorana and Dirac masses for the right-handed neutrinos respectively. 
 \section{The SD approach to condensate formation in the Higgs-Yukawa system}
In this section we inquire into the conditions for which fermion bilinear condensates can get formed. As mentioned in \cite{HungXiong}, for an initial value of the Yukawa coupling which is large enough, the heavy fourth generations condense through the exchange of a fundamental Higgs doublet when the Yukawa coupling exceeds a certain critical value. The electroweak symmetry breaking is then driven by such a fermion bilinear condensate. By applying the apparatus developed in \cite{HungXiong} to the EW-scale $\nu_R$ model, the critical Yukawa couplings will be derived.\par
Let us recall that the EW-scale $\nu_R$ model is spontaneously broken by the VEVs of both Higgs triplets and doublets \cite{HungMH}, \cite{Th.Hung}. We consider two types of condensates here: that which is generated by the exchange of the fundamental Higgs triplet $\tilde{\chi}$ between two right-handed neutrinos and the other which is generated by the exchange of the fundamental Higgs douplet $\Phi_{2M}$ between two mirror quarks.
To preserve custodial $SU(2)_D$ symmetry, one assumes that non-zero expectation values for quark bilinears in the EW-scale $\nu_R$ model satisfy the conditions: $\langle\bar U_L^MU_R^M\rangle=\langle\bar D_L^MD_R^M\rangle $. This condition is satisfied by assuming $g_{u^M}=g_{d^M}=g_{q^M}$.\par

It is common to study the evolution of Yukawa couplings by investigating their $\beta$ functions. We restrict our discussion to the study of energy scale at which Landau poles appear, the $\beta$ functions of  Yukawa couplings are then calculated up to one loop order. These functions together with the critical Yukawa couplings given by SD equations define the scale of condensation, that is to say, the electroweak symmetry breaking scale. At that scale, our model will gain the additional composite Higgs, namely, composite Higgs doublet and composite Higgs triplet formed from right-handed neutrino and mirror fermions. This issue will be shown in the next section. In the following analysis we have made the following assumptions  
\begin{itemize}
\item It will be assumed in our discussion of condensate formation that the fundamental Higgs fields are massless. These fields then have no VEV at tree level. 

\item Triplet condensates form from $\nu_R$ condensate; doublet condensates form from $q^M$ condensate; no other condensates; $\phi_2^0$ and $\xi^0$ develop VEVs through couplings (see below).

\end{itemize}
Let us begin by considering the Yukawa Lagrangians given by Eqs.~\eqref{mm2} and \eqref{mm3} which are directly related to condensates.
\begin{figure}[ht]

\begin{center}

\includegraphics[scale=0.17]{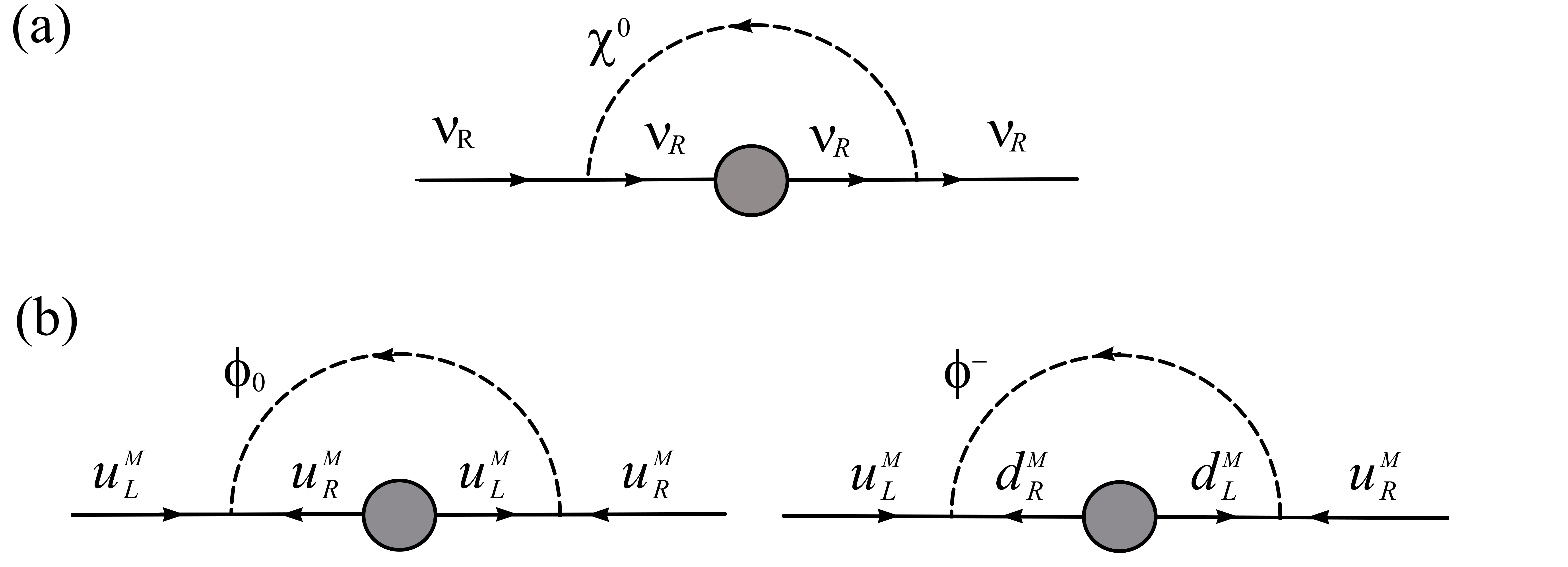}\\

\end{center}

\caption{Graph contributing to the right-hand side of the SD equation 
for the (a) right-handed neutrino
self-energy $\Sigma_{\nu_R}$, (b) mirror quark self-energy $\Sigma_{q^M}$. }\label{trip}

\end{figure}
The SD equations for the right-handed neutrino $\nu_R$ and mirror quark $q^M$ self-energy  which relevant Feynman diagrams drawn in Fig.~\ref{trip} read  
\begin{eqnarray}
\label{SEnu}
\Sigma_{\nu_R}(p)&=&\frac{g_M^2}{(2\pi)^4}\int d^4q\frac{1}{(p-q)^2}\frac{\Sigma_{\nu_R}(q)}{q^2+\Sigma^2_{\nu_R}(q)},\\
\label{SEq}
\Sigma_{q^M}(p)&=&2\times\frac{ g_{q^M}^{2}}{(2\pi)^4}\int d^4q\frac{1}{(p-q)^2}\frac{\Sigma_{q^M}(q)}{q^2+\Sigma^{2}_{q^M}(q)}.
\end{eqnarray}
By converting Eqs.~\eqref{SEnu}, \eqref{SEq} to differential equations and setting $\alpha_{\nu_R}=\frac{g_M^2}{4\pi}$, $\alpha_{q^M}=\frac{g_{q^M}^{2}}{4\pi}$, Eqs.~\eqref{SEnu} and \eqref{SEq} become
\begin{eqnarray}
\label{SEnu1}
\Box \Sigma_{\nu_R}(p)&=&-\left(\frac{\alpha_{\nu_R}}{\alpha_{\nu_R}^c}\right)\frac{\Sigma_{\nu_R}(p)}{p^2+\Sigma^2_{\nu_R}(p)},\\\label{SEq1}\Box \Sigma_{q^M}(p)&=&-\left(\frac{\alpha_{q^M}}{\alpha_{q^M}^c}\right)\frac{\Sigma_{q^M}(p)}{p^2+\Sigma^2_{q^M}(p)},
\end{eqnarray}
with the following boundary conditions 
\begin{eqnarray}
\label{SEnu2}
&&\begin{cases}\lim_{p\rightarrow 0}p^4\frac{d\Sigma_{\nu_R}(p)}{dp^2}=0\\ \lim_{p\rightarrow\Lambda}p^2\frac{d\Sigma_{\nu_R}(p)}{dp^2}+\Sigma_{\nu_R}(p)=0\end{cases},\\&&\begin{cases}\lim_{p\rightarrow 0}p^4\frac{d\Sigma_{q^M}(p)}{dp^2}=0\\ \lim_{p\rightarrow \Lambda}p^2\frac{d\Sigma_{q^M}(p)}{dp^2}+\Sigma_{q^M}(p)=0\end{cases},
\end{eqnarray}
and the critical Yukawa couplings of right-handed neutrino and mirror quark are  
\begin{eqnarray}
\label{ccnu}\alpha_{\nu_R}^c&=&\pi,\\\label{ccq}
\alpha_{q^M}^c&=&\frac \pi 2,
\end{eqnarray}
where we have assumed that $\Sigma_{u^M}(p)=\Sigma_{d^M}(p)=\Sigma_{q^M}(p)$.
The numerical and analytic solutions to these differential equations are similar to those proposed in \cite{HungXiong}. The values given in Eqs.~\eqref{ccnu} and \eqref{ccq} are called critical values since in \cite{Leung} there are two classes of asymptotic solutions at large momentum for different values of $\alpha$. Above these critical values, the solutions to these SD equations for mirror fermions have the form of bound states.  

Next, the quantity which is closely related to the dynamical breaking of the EW symmetry is the condensates of the mirror quarks and right-handed neutrinos. We have
\begin{eqnarray}
\label{cq1}
\langle{\bar u_L^M u_R^M}\rangle&=&\langle{\bar d_L^M d_R^M}\rangle\nonumber\\
&\approx&-\frac 3 {\pi^2}\left(\frac{\alpha_{u^M}^c}{\alpha_{u^M}}\right)\Lambda\Sigma^2_{u^M}(0)\sin\left[\sqrt{\frac{\alpha_{u^M}}{\alpha_{u^M}^c}-1}\right].
\end{eqnarray}
Likewise, the condensate state of right-handed neutrino would be
\begin{eqnarray}
\label{cq2}
\langle{\nu_R^{T}\sigma_2 \nu_R}\rangle&\approx &-\frac 1 {\pi^2}\left(\frac{\alpha_{\nu_R}^c}{\alpha_{\nu_R}}\right)\Lambda\Sigma^2_{\nu_R}(0)\sin\left[\sqrt{\frac{\alpha_{\nu_R}}{\alpha_{\nu_R}^c}-1}\right]
\end{eqnarray}
As mentioned in \cite{HungXiong}, the cut off $\Lambda$ in Eqs.~\eqref{cq1} and \eqref{cq2} could be as low as $\mathcal O(\mbox{TeV})$ in order to avoid the fine tuning scenario. Therefore, these condensates are directly related to $v_\chi$ and $v_{\Phi_{2M}}$, namely
\begin{eqnarray}\label{cnu}
\langle{\nu_R^{T}\sigma_2 \nu_R}\rangle&\sim & \mathcal O(-v_{\chi}^3),\\\label{cqm}\langle{\bar q_L^M q_R^M}\rangle&\sim & \mathcal O(-v_{\Phi_{2M}}^3),
\end{eqnarray}
 where it will be seen later that $v_\chi$, $v_{\Phi_{2M}}$ are the induced VEV of $\chi$, $\Phi_{2M}$, respectively. It will shown below that $v_\chi$ and $v_{\Phi_{2M}}$ are of the order of EW scale then no fine tuning is required if $\Lambda\sim\mathcal O(1~\mbox{TeV})$.
 \section{One-loop $\beta$ functions for Yukawa couplings in the EW-scale $\nu_R$ model}
As the previous section pointed out, to get condensates with the absence of fine tuning problem, the cut-off $\Lambda$ of new physics should be of the order of $\mathcal O(1~ TeV)$. Hence, the one-loop $\beta$ function will be calculated in this section in order to find the appropriate initial Yukawa couplings from which condensates form and the Landau poles appear at that scale. 

The one-loop contributions to the Yukawa coupling renormalization constant are composed of four terms: vertex corrections, fermion self-energy, scalar self-energy and gauge interactions. At high energies, the gauge couplings are small compared with the Yukawa couplings and their contribution to the $\beta$-function will be neglected. For this reason, to compute one-loop $\beta$ function we need only consider the first three terms. The evolution of Yukawa couplings of right-handed neutrinos and mirror fermions in the EW-scale $\nu_R$ model \cite{HungMH} are given by
\begin{eqnarray}
\label{betanu}
16\pi^2\frac {dg_M}{dt}&=&\frac {13} 2g_M^3+\frac 1 2\left(g_{e^{M}}\right)^2g_M,\\\label{betae}16\pi^2\frac {dg_{e^M}}{dt}&=&\frac {11}2\left(g_{e^{M}}\right)^3+\frac 3 4g_M^2g_{e^M},\\\label{betaq}16\pi^2\frac {dg_{q^M}}{dt}&=&6\left(g_{q^{M}}\right)^3.
\end{eqnarray}
These Renormalization Group Equations (RGEs) can be solved numerically \cite{HungLe} and the results depend on initial Yukawa couplings. However, one may query the implications of the initial Yukawa couplings chosen once solutions to RGEs was found. Why some observable quantities, such as masses were not used as initial values in this situation. The answer would rely on our current paper's hypothesis that matter acquires no mass until DEWSB. Therefore, these initial Yukawa couplings would infer some physical quantities from a different point of view where EWSB occurs, say, the \textit{naive} masses \cite{HungXiong}. From now, the solution to RGEs can be investigated by using initial values of \textit{naive} mirror fermion masses in the EW-scale $\nu_R$ model.  
\begin{figure}[h]  
\begin{center}
\includegraphics[scale=0.40]{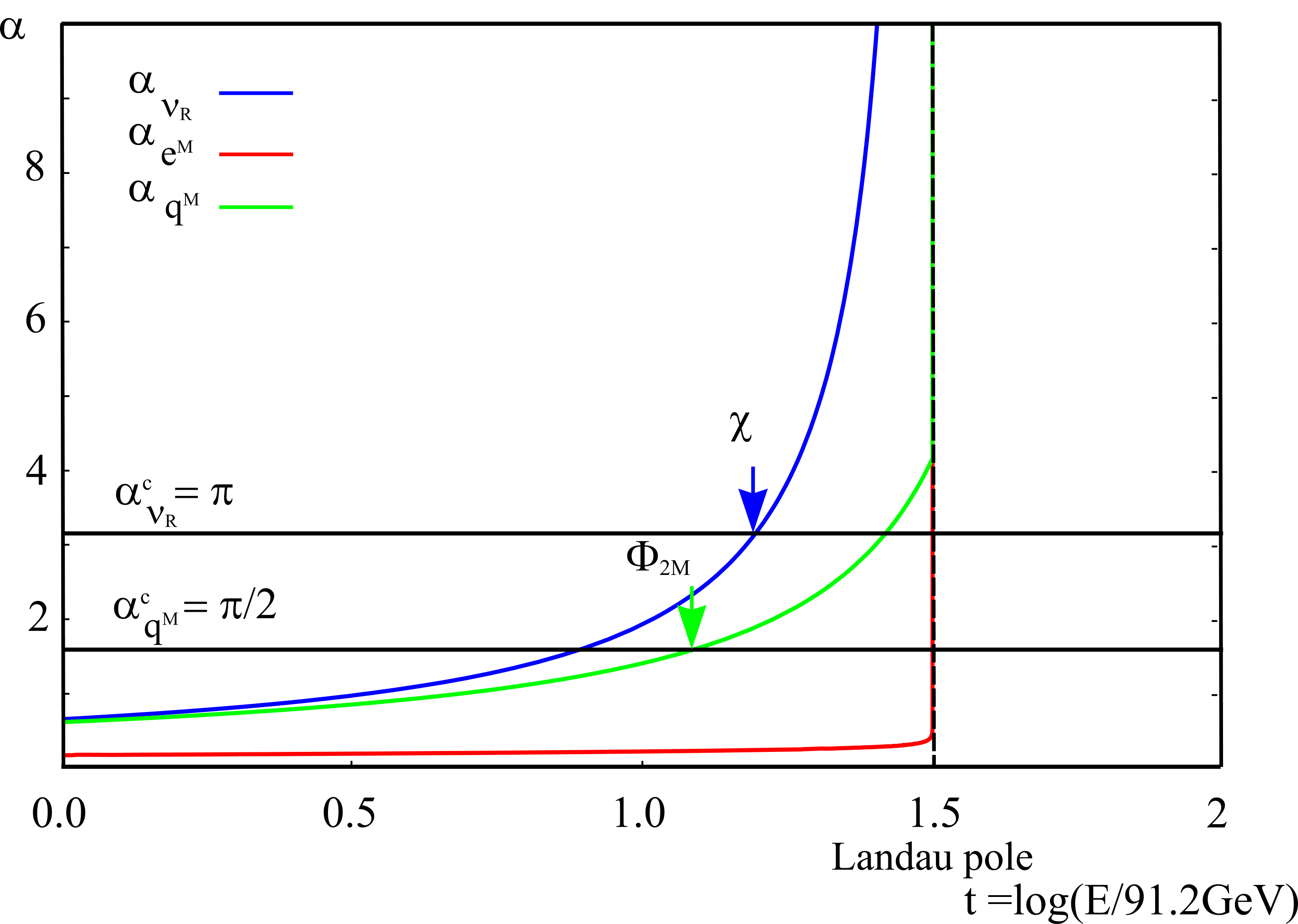}
 \caption{The evolution of Yukawa couplings where the initial masses  of $\nu_R, e^M$ and $q^M$ are $200 ~\mbox{GeV},102~\mbox{GeV}$ and $202~\mbox{GeV}$, respectively. The blue and green arrows indicate the energy values where the Higgs triplet $\chi$ and Higgs doublet $\Phi_{2M}$ correspondingly get VEVs.} \label{noTop.jpg}
 \end{center}
 \end{figure}
\par The bounds on masses of mirror fermions was presented in \cite{Th.Hung} using characteristics of the 2012 discovery of LHC of  the 125 GeV boson. This boson was assumed to exist in two states: \textit{Dr. Jekyll} or \textit{Mr. Hyde} depending on the influence of neutral Higgs $H_1^0$. While in \textit{Dr. Jekyll} state in which $H_1^0$ as a dominant component tend to behave more like the SM Higgs boson, the \textit{Mr. Hyde} state with $H_1^0$ as a sub-dominant component is very different. The upper bounds on masses of fermions in the EW-scale $\nu_R$ model in scenarios \textit{Dr. Jekyll} and \textit{Mr. Hyde} are $120$ GeV and $700$ GeV, respectively. \par The solution to RGEs can be investigated by using proper initial values of \textit{naive} mirror fermion masses in the EW-scale $\nu_R$ model. These values are satisfied the case as initially pointed out, where triplet and doublet condensates form from $\nu_R$ and $q^M$ condensates. In particular, Fig.~\ref{noTop.jpg} shows the situation where the initial masses  of $\nu_R, e^M$ and $q^M$ are $200 ~\mbox{GeV},102~\mbox{GeV}$ and $202~\mbox{GeV}$, respectively. As shown in Fig.~\ref{noTop.jpg}, the Yukawa couplings  increase dramatically as energy increases and Ladau pole singularities appear at $t=1.50$ ($E=2.89~\mbox{TeV}$). By using the critical values of Yukawa couplings found above, we immediately find that the values of $t$ at which condensates of right-handed neutrino and mirror quark arise are estimated to be $1.19$ and $1.09$, respectively, i.e. at the order of $\mathcal O(1~\mbox{TeV})$. Then the Higgs fields $\chi,\Phi_{2M}$ will get VEV through condensations of fermions.
\section{DEWSB in the EW-scale $\nu_R$ model}
In this section we present how  electroweak symmetry is dynamically broken. To study DEWSB, the elementary scalar fields in the EW-scale $\nu_R$ model are assumed to have no VEVs at tree level \cite{HungXiong}. The scale-invariant potential for fundamental Higgs can be written as
\begin{eqnarray}
\label{funHiggs}
V_f&=&V_f(\Phi_2,\Phi_{2M},\chi)=\lambda_1\left[Tr\Phi_2^+\Phi_2\right]^2+\lambda_2\left[Tr\Phi_{2M}^+\Phi_{2M}\right]^2+\lambda_3\left[Tr\chi^+\chi\right]^2\nonumber\\&&+\lambda_4\left[
Tr\Phi_2^+\Phi_2+Tr\Phi_{2M}^+\Phi_{2M}+Tr\chi^+\chi\right]^2\nonumber\\&&+\lambda_5\left[\left(Tr\Phi_2^+\Phi_2\right)\left(Tr\chi^+\chi\right)-2\left(Tr\Phi_2^+\frac{\tau^a}{2}\Phi_2\frac{\tau^b}{2}\right)\left(Tr\chi^+T^a\chi T^b\right)\right]\nonumber\\&&+\lambda_6\left[\left(Tr\Phi_{2M}^+\Phi_{2M}\right)\left(Tr\chi^+\chi\right)-2\left(Tr\Phi_{2M}^+\frac{\tau^a}{2}\Phi_{2M}\frac{\tau^b}{2}\right)\left(Tr\chi^+T^a\chi T^b\right)\right]\nonumber\\&&+\lambda_7\left[\left(Tr\Phi_2^+\Phi_2\right)\left(Tr\Phi_{2M}^+\Phi_{2M}\right)-\left(Tr\Phi_2^+\Phi_{2M}\right)\left(Tr\Phi_{2M}^+\Phi_{2}\right)\right]\nonumber\\&&+\lambda_8\left[3Tr\chi^+\chi\chi^+\chi-\left(Tr\chi^+\chi\right)^2\right],
\end{eqnarray}
from which the Higgs fields have no mass terms. This assumption is,  however, invalid at low energy. Under the condensate scale, the fundamental Higgs fields  would become the composite Higgs fields. Let us consider the case mentioned in the previous section where Higgs triplet and Higgs doublet condensates form from $\nu_R$ and $q^M_R$ condensates. The Higgs effective potential can be written in terms of fundamental Higgs and the additional composite Higgs $\Phi_{2Mc} $ and $ \widetilde\chi_c$, in particular 
\begin{eqnarray} 
\label{Higgseff}
V_{eff}=V_{eff}(\Phi_2,\Phi_{2M},\widetilde\chi,\xi,\Phi_{2Mc},\widetilde\chi_c).
\end{eqnarray} 
Instead of investigating the explicit expression of the Higgs effective potential, the issue of mass generation here would be the main focus. It turns out that the potential given in Eq.~\eqref{Higgseff} is no longer scale-invariant and the Higgs fields will acquire masses due to the appearance of condensates. The contributions of induced $\mu^2$ of $\chi^0$ and $\phi_{2M}^0$ to the Higgs effective potential 
come from the dimension 5 operators involving the neutral fundamental scalar $\chi^{0}$ and $ \phi_{2M}^0$ 
\begin{eqnarray}\label{chi0}
\frac 1 2\left\{\frac{g_M^2}{\Sigma_{\nu_R}(0)}\left(\nu_R^{T}\sigma_2\nu_R \right)\right\}\left|\chi^{0}\right|^2,\\\label{phi2M0}
\frac 1 2\left\{\frac{g_{q^M}^2}{\Sigma_{q^M}(0)}\left(\bar u^M_Lu^M_R+\bar d^M_Ld^M_R\right)\right\}\left|\phi_{2M}^0\right|^2,
\end{eqnarray}
where $\Sigma_{\nu_R}(0)$, $\Sigma_{q^M}(0)$ are the dynamical masses of right-handed neutrino and mirror quark, respectively. 
When condensates get formed, these terms will give rise to the following negative effective mass squared terms for $\chi^{0}$ and $\phi_{2M}^0$ as follows
\begin{eqnarray}\label{chi01}
\frac 1 2\frac{g_M^2}{\Sigma_{\nu_R}(0)}\langle{\nu_R^{T}\sigma_2\nu_R }\rangle\left|\chi^{0}\right|^2,\\\label{phi201}
\frac{g_{q^M}^2}{\Sigma_{q^M}(0)}\langle{\bar u^M_Lu^M_R}\rangle\left|\phi_{2M}^0\right|^2,
\end{eqnarray}
\begin{figure}[ht]

\begin{center}

\includegraphics[scale=0.21]{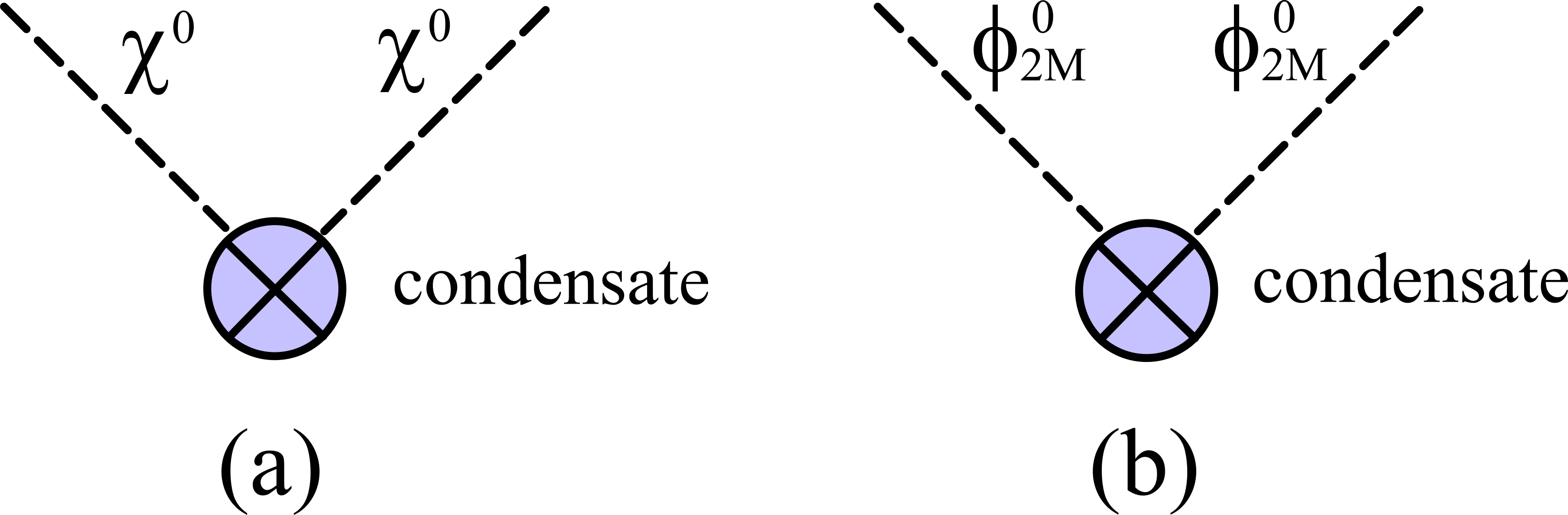}\\

\end{center}

\caption{Diagram giving mass to (a) $\chi^0$, (b) $\phi_{2M}^0$. }\label{tripp}

\end{figure}
which are illustrated in Fig.~\ref{tripp}. \par
\begin{figure}[ht]

\begin{center}

\includegraphics[scale=0.21]{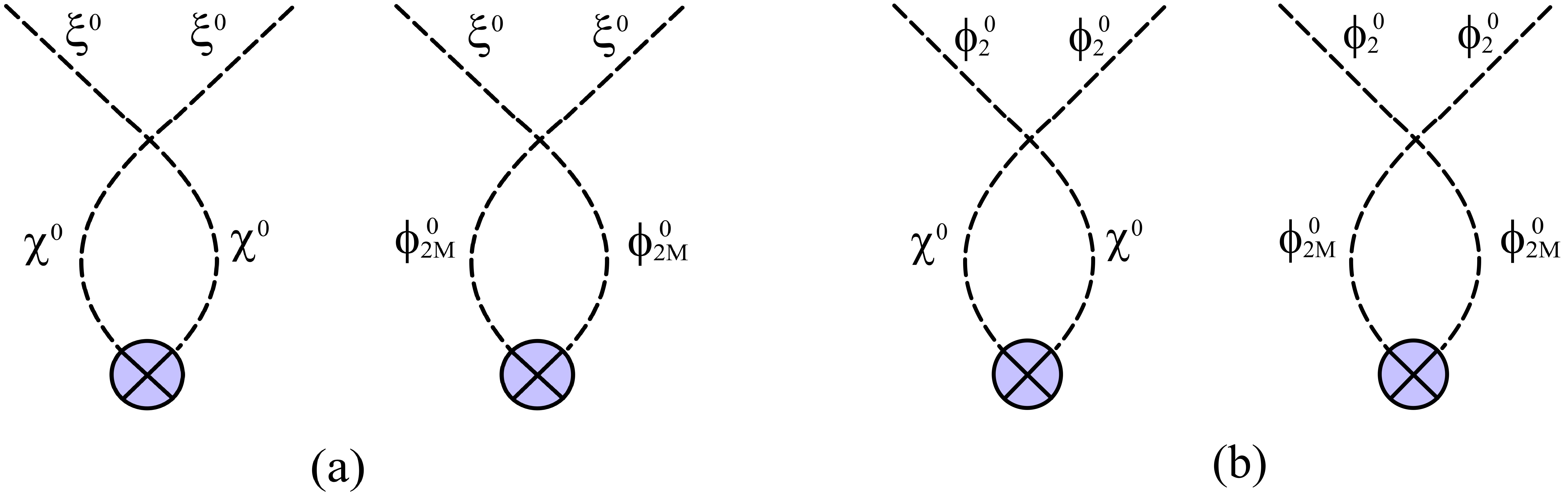}\\

\end{center}

\caption{Diagram giving mass to (a) $\xi^0$, (b) $\phi_2^0$.}\label{xiphi2mass.jpg}

\end{figure}

 
In the case of the fundamental Higgs $\xi$ which has no interaction with fermions, a mass term for $\xi^0$ can be obtained through quadratic interactions with $\chi^0$ and $ \phi_{2M}^0$. The corresponding Feynman diagrams  are shown in Fig.~\ref{xiphi2mass.jpg}a. and contributions of induced $\mu^2$ of $\xi^0$ to Higgs effective potential involving the neutral fundamental scalar $\xi^{0}$ are as follows
\begin{eqnarray}\label{xixi}
\frac 1 2\Bigg\{\frac {g_M^2}{\Sigma_{\nu_R}(0)}\langle{\nu_R^{T}\sigma_2\nu_R }\rangle I_\chi^{(1)}+\frac{2g_{q^M}^2}{\Sigma_{q^M}(0)}\langle{\bar u^M_Lu^M_R}\rangle I_{\phi_{2M}}^{(1)}\Bigg\}\left|\xi^0\right|^2,
\end{eqnarray}
where the integrals $I_\chi^{(1)}$ and $I_{\phi_{2M}}^{(1)}$ are given by 
\begin{eqnarray}\label{Ichi1}
I_\chi^{(1)}&=&\frac {4\lambda_3+\lambda_4-\lambda_8}{16\pi^2}\ln\left(\frac {\Lambda^2}{m^2}\right),\\\label{Iphi2m1}
I_{\phi_{2M}}^{(1)}&=&\frac {4\lambda_4+2\lambda_6}{16\pi^2}\ln\left(\frac {\Lambda^2}{m^2}\right),\end{eqnarray}
with $\Lambda$ being a cutoff proportional to the condensate scale and $m$ being a renormalization scale.
Similarly, Feynman diagrams generating mass for $\phi_2^0$ are shown in Fig.~\ref{xiphi2mass.jpg}b and one has the same form for $\left|\phi_2^0\right|^2$ term as follow  
\begin{eqnarray}\label{phi22}
\frac 1 2\Bigg\{\frac {g_M^2}{\Sigma_{\nu_R}(0)}\langle{\nu_R^{T}\sigma_2\nu_R }\rangle I_\chi^{(2)}+\frac{2g_{q^M}^2}{\Sigma_{q^M}(0)}\langle{\bar u^M_Lu^M_R}\rangle I_{\phi_{2M}}^{(2)}\Bigg\}\left|\phi_2^0\right|^2,
\end{eqnarray}
where the integrals $I_\chi^{(2)}$ and $I_{\phi_{2M}}^{(2)}$ are given by
\begin{eqnarray}\label{Ichi1}
I_\chi^{(2)}&=&\frac {8\lambda_4-4\lambda_5}{16\pi^2}\ln\left(\frac {\Lambda^2}{m^2}\right),\\\label{Iphi2m1}
I_{\phi_{2M}}^{(2)}&=&\frac {8\lambda_4+2\lambda_7}{16\pi^2}\ln\left(\frac {\Lambda^2}{m^2}\right).\end{eqnarray}
One can see that the VEV's vanish when the condensates vanish. The presence of the terms in Eqs.~\eqref{chi01}, \eqref{phi201}, \eqref{xixi} and \eqref{phi22} enables the fundamental Higgs fields to get non-zero VEVs. The VEVs of $\chi,\xi,\Phi_2$ and $\Phi_{2M}$ are assumed to be $v_\chi,v_\xi,v_{\Phi_2}$ and $v_{\Phi_{2M}}$, respectively.
When $\chi$, $\Phi_2$ and $\Phi_{2M}$ get VEVs
\begin{eqnarray}
\label{chivev}
\langle\chi\rangle&=&\left(\begin{matrix}
v_\chi& 0 & 0\\0 & v_
\xi & 0\\ 0& 0& v_\chi
\end{matrix}\right),\\\langle\Phi_{2}\rangle&=&\left(\begin{matrix}
v_{\Phi_2}/\sqrt 2& 0\\ 0& v_{\Phi_2}/\sqrt 2
\end{matrix}\right),
\end{eqnarray}
and
\begin{equation}
\label{phi2vev}
\langle\Phi_{2M}\rangle=\left(\begin{matrix}
v_{\Phi_{2M}}/\sqrt 2& 0\\ 0& v_{\Phi_{2M}}/\sqrt 2
\end{matrix}\right),
\end{equation}
the symmetry is dynamically broken  from global $SU(2)_L\otimes SU(2)_R$ down to the custodial $ SU(2)_D$. A detailed discussion of the minimization of the Higgs effective potential is in progress.
At tree level, the gauge boson masses are obtained by kinetic part of the Higgs Lagrangian
\begin{eqnarray}
\label{kinHiggs}
\mathcal L_{kin}&=&\frac 1 2\left|\partial_\mu \phi_S\right|^2+\frac 1 2Tr\left[\left(D_\mu\Phi_{2M}\right)^+\left(D^\mu\Phi_{2M}\right)\right]+\frac 1 2Tr\left[\left(D_\mu\Phi_2\right)^+\left(D^\mu\Phi_2\right)\right]\nonumber\\&&+\frac 1 2Tr\left[\left(D_\mu\chi\right)^+\left(D^\mu\chi\right)\right],
\end{eqnarray} 
where
\begin{eqnarray}
\label{daohamphi1}
D_\mu\Phi_2&\equiv&\partial_\mu\Phi_2+i\frac g2\left(W\cdot \tau\right)\Phi_2-i\frac{g'}2\Phi_2B\tau_3,\\\label{daohamphi2}
D_\mu\Phi_{2M}&\equiv&\partial_\mu\Phi_{2M}+i\frac g2\left(W\cdot \tau\right)\Phi_{2M}-i\frac{g'}2\Phi_{2M}B\tau_3,\\\label{daohamchi}
D_\mu\chi &\equiv&\partial_\mu\chi+ig\left(W\cdot T\right)\chi-ig'\chi BT^3,
\end{eqnarray}
and $\tau_i/2$, $T^i$ are the $2\times 2$ and $3\times 3$ representation matrices of $SU(2)$ respectively. One gets
\begin{eqnarray}\label{mWH}
m_W^2&=&\frac{g^2 v^2}{4}=\frac{g^2}{4}\left(v_{\Phi_2}^2+v_{\Phi_{2M}}^2+4v_\xi^2+4v_\chi^2\right),\\\label{mZH}
m_Z^2&=&\frac{g^2}{4\cos^2\theta_W}\left(v_{\Phi_2}^2+v_{\Phi_{2M}}^2+4v_\xi^2+4v_\chi^2\right).
\end{eqnarray}
At tree level, $\rho=1$, 
when proper alignment of the two VEVs is assured, we obtained 
\begin{equation}\label{xichi}v_{\xi}=v_\chi.\end{equation} 
In order to satisfy Eq.~\eqref{xichi}, one puts a constraint on the parameters appearing in Eq.~\eqref{xixi}. This is so in order to guarantee that we have a custodial symmetry. In this situation, 
\begin{equation}
\label{vew}
v^2=v_{\Phi_2}^2+v_{\Phi_{2M}}^2+8v_\chi^2.
\end{equation}
\section{VEV of the Higgs singlet $\phi_S$}
When condensate states get formed, the fundamental Higgs singlet $\phi_S$ simultaneously develops VEV which gives a Dirac mass to the neutrinos. As discussed in \cite{HungMH}, the Dirac neutrino mass comes from the
Lagrangian given in Eq.~\eqref{dontuyenY1}. With $\left\langle\phi_S\right\rangle=v_S$, one obtains a Dirac neutrino mass $m_D=g_{Sl}v_S$. As in \cite{HungMH}, since the light neutrino masses $\sim m_D^2/M_R$ are constrained to be $<O(\mbox{eV})$ and $M_R \sim O(\Lambda_{EW})$, it follows that $m_D = g_{Sl} v_S < O(10^5~\mbox{eV})$. From \cite{neulepmirror2}, $g_{Sl}$ is constrained to be less than $10^{-3}$ using the present upper bound on the rate of $\mu \rightarrow e \gamma$. This implies that $v_S <100~ \mbox{MeV} < \Lambda_{EW}$. We show below that the induced $v_S$ is indeed so.\par
\begin{figure}[ht]

\begin{center}

\includegraphics[scale=0.2]{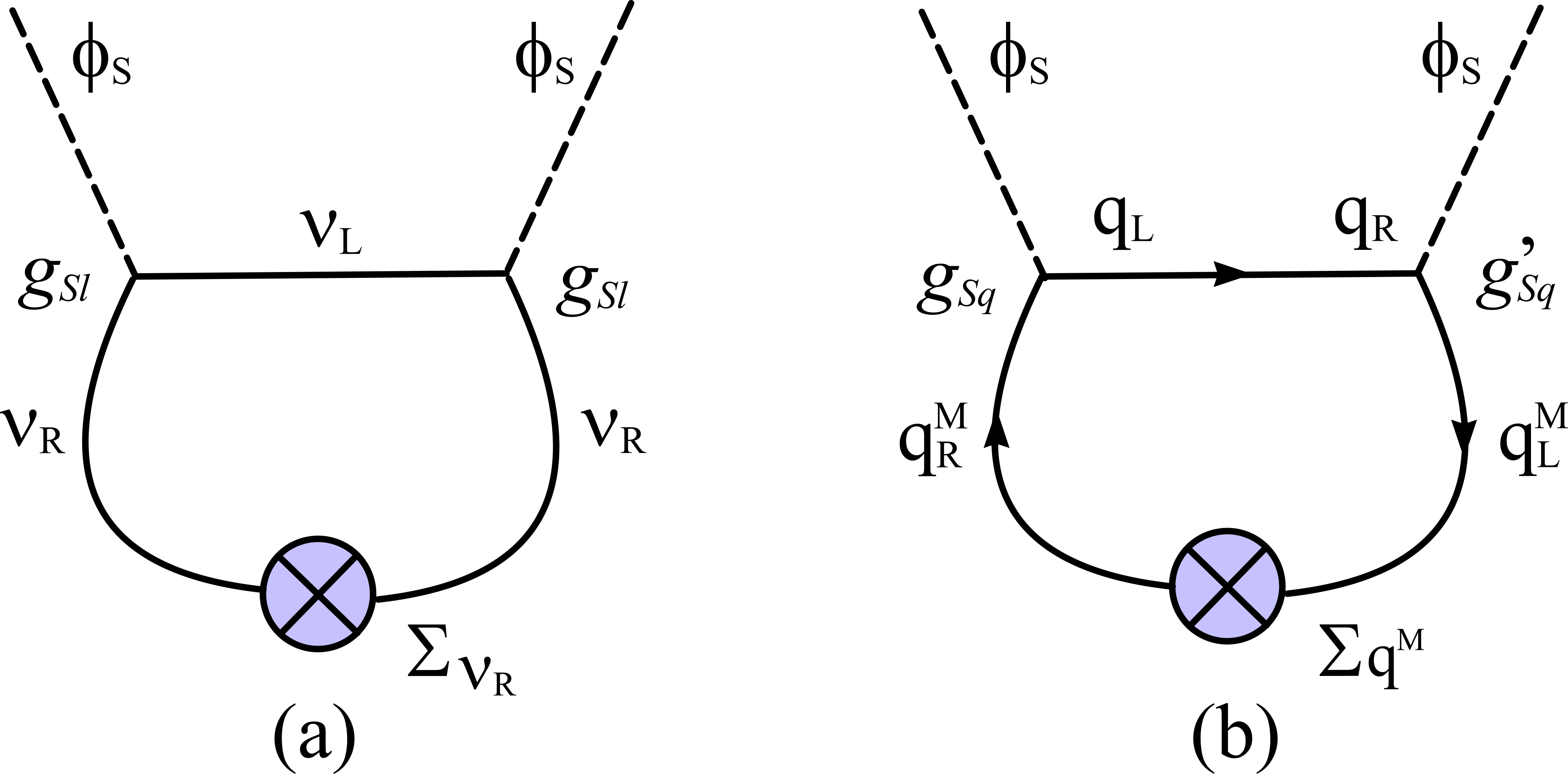}\\

\end{center}

\caption{Diagram giving VEV to $\phi_S$: (a) from right-handed neutrino self-energy, (b) from mirror quark self-energy.}\label{phiS.jpg}

\end{figure}

 
After DEWSB occurs, the VEV of $\phi_S$ could be roughly estimated through condensate states. Feynman diagrams in Fig.~\ref{phiS.jpg} show how $\phi_S$ develops VEV.  
Let us first consider Fig.~\ref{phiS.jpg}a. The right-handed neutrino self-energy $\Sigma_{\nu_R}$ would be proportional to the condensate scale $\Lambda_{cond}$, $\Sigma_{\nu_R} \sim \Lambda_{cond}$. The $\left|\phi_S\right|^2$ term then has the following form
\begin{equation}
\label{11}
\frac {g_{Sl}^2}{16\pi^2}\Lambda\Lambda_{cond}\left|\phi_S\right|^2\sim\frac {g_{Sl}^2}{16\pi^2}\Lambda_{cond}^2\left|\phi_S\right|^2,
\end{equation}
where the momentum cutoff is $\Lambda \sim\Lambda_{cond}$. 
Since from recent studies of $\mu\rightarrow e\gamma$ \cite {neulepmirror2}, $g_{Sl}\leq 10^{-3}$ and more like $10^{-5}$, then one obtains
\begin{equation}
\label{11}
\frac {g_{Sl}^2}{16\pi^2}\Lambda_{cond}^2\left|\phi_S\right|^2\sim 10^{-12}\Lambda_{cond}^2\left|\phi_S\right|^2 .
\end{equation} 
The scale of the $\phi_S$-VEV will then be expected to be proportional to $10^{-6}\Lambda_{cond}$ which is roughly one wants. Therefore, the induced VEV of $\phi_S$ comes out "naturally" small. Similarly, one could get the same order of induced VEV of $\phi_S$ coming from Fig.~\ref{phiS.jpg}b if we assume $g_{Sq}\sim g_{Sq}'\sim g_{Sl}$. \par
The induced VEV of $\phi_S$ emerges "naturally" small as a result of the constraint $g_{Sl}\leq 10^{-3}$ studied in \cite{neulepmirror2}. Turning this argument around, the smallness of neutrino masses implies that $g_{Sl}$ has to small within the present framework. The implication of this result will be presented elsewhere.
\section{Conclusions}
As with the 4th-generation scenario of \cite{HungXiong}, we started out in this paper with the EW-scale $\nu_R$ model with massless scalars at tree level. The Higgs fields will become composite Higgs when fermions in the EW-scale $\nu_R$ model condense. Condensates and critical values were dynamically investigated within the framework of SD approach. If initial values of the Yukawa coupling exceed corresponding critical values, the condensation of right-handed neutrino and mirror fermions through the exchange of fundamental Higgs occur. By studying of the evolution of one-loop $\beta$ functions of Yukawa couplings, we have also shown that the energy where condensate states form is at the the order of $\sim\mathcal O(1~\mbox{TeV})$. The electroweak symmetry breaking is then driven by fermion bilinear condensates at that scale. In the present work, the case where condensates of $\nu_R$ and $q^M$ give rise to composite Higgs $\widetilde\chi_c$ and $\Phi_{2Mc}$ was considered. After the formation of right-handed neutrino and mirror quarks condensates, Higgs triplet $\chi$ and Higgs doublet $\phi_{2M}$ will develop VEVs through induced $\mu^2$ terms. The others, $\phi_2$ and $\xi$ acquire masses because of quadratic couplings with $\chi^0$ and $ \phi_{2M}^0$. The global $SU(2)_L\otimes SU(2)_R$ symmetry is then dynamically broken down to the custodial $ SU(2)_D$.\par This study actually provides an interesting result. First, a model of DEWSB was provided where $\nu_R$'s and mirror quarks obtained dynamical masses proportional to $\Lambda_{EW}$. The other interesting feature of this work is that within the framework of DEWSB, the induced VEV of $\phi_S$ would be naturally smaller than the electroweak scale because of the smallness of $g_{Sl}$. This implies that $m_D \ll M_R$ and the light neutrinos are naturally light. Last but not least, there is also a recent interest concerning the possibility that the Higgs boson is a composite of neutrinos \cite{hill}.\par 
The model presented in this paper is scale-invariant at tree-level above the condensate scale. Scale invariance is broken at "low energy" and it is an interesting question to see if there are any phenomenological consequences. This question is under investigation.
\section*{Acknowledgments}

NNL would like to thank the Department of Physics, University of Virginia for hospitality during the earlier stages of this investigation. Financial support of Hue University is gratefully acknowledged.


\begin{thebibliography}{0}
\bibitem{models}
The list of references related to these topics is extensive and it is impossible to put it in this note.
\end{thebibliography}
\end{document}